\documentclass[usenatbib,twocolumn]{mn2e}
\usepackage{epsfig}
\usepackage{amsmath}
\usepackage{lscape}
\usepackage{longtable}
\usepackage{anysize}
\usepackage{multirow}
\def\beq{\begin{equation}}
\def\eeq{\end{equation}}
\def\bey{\begin{eqnarray}}
\def\eey{\end{eqnarray}}

\def\Msun{\,{\rm M_\odot}}
\def\Msunh{\, h^{-1}{\rm M_\odot}}

\def\gs{\mathrel{\raise1.16pt\hbox{$>$}\kern-7.0pt
\lower3.06pt\hbox{{$\scriptstyle \sim$}}}}
\def\ls{\mathrel{\raise1.16pt\hbox{$<$}\kern-7.0pt
\lower3.06pt\hbox{{$\scriptstyle \sim$}}}}
\def\gtsima{$\; \buildrel > \over \sim \;$}
\def\ltsima{$\; \buildrel < \over \sim \;$}
\def\prosima{$\; \buildrel \propto \over \sim \;$}
\def\gsim{\lower.5ex\hbox{\gtsima}}
\def\lsim{\lower.5ex\hbox{\ltsima}}
\def\simgt{\lower.5ex\hbox{\gtsima}}
\def\simlt{\lower.5ex\hbox{\ltsima}}
\def\simpr{\lower.5ex\hbox{\prosima}}
\def\la{\lsim}
\def\ga{\gsim}

\usepackage{color}

\marginsize{1.0in}{1.0in}{1.0in}{1.0in}

\begin{document}

\title[Quenching of Satellite Galaxies]
{Star Formation and Quenching of Satellite Galaxies}
\author[Zhankui Lu and H.J. Mo]
   {\parbox[t]{\textwidth}{
       Zhankui Lu$^{1}$\thanks{E-mail: lv@astro.umass.edu},
       H.J. Mo$^{1}$
}\\
$^1$Department of Astronomy, University of Massachusetts, Amherst MA 01003-9305, USA}

\date{Accepted ........ Received .......; in original form ......}
\pubyear{2014}

\maketitle 

\label{firstpage}

\begin{abstract}
We study the quenching of satellite galaxies by gradual depletion of gas 
due to star formation, by ram-pressure striping and by tidally triggered 
starburst. Using progenitors constrained by the empirical model of Lu et al.,
in which outflow loading factor is low, we do not find an over-quenching 
problem in satellites even if there is no further cold gas supply from the 
cooling of the halo gas after a galaxy is accreted by its host. Gradual 
depletion alone predicts a unimodal distribution in specific star formation, 
in contrast to the bimodal distribution observed, and under-predicts the 
quenched fraction in low mass halos. Ram-pressure stripping nicely 
reproduces the bimodal distribution but under-predicts the quenched fraction 
in low-mass halos. Starbursts in gas-rich satellites triggered by tidal 
interactions with central galaxies can nicely reproduce the quenched 
satellite population in low-mass halos, but become unimportant for            
low-mass satellites in massive halos. The combined processes, together 
with the constrained progenitors, can reproduce the observed 
star formation properties of satellites in halos of different masses.              
\end{abstract}



\section{Introduction}
\label{intro}

The impact of the intra-halo environments on the 
star formation in satellite galaxies has been noticed for 
decades \citep[see \S15.5 in][]{Mo10}, and has been investigated in detail
using groups/clusters of galaxies.  
For example, using the group catalogue 
constructed from SDSS DR2, \citet{Weinmann06a} examined
how the fractions of late and early type satellite galaxies 
change with galaxy luminosity and the mass of the halo that host 
the satellites, and found that the early type fraction increases 
strongly with halo mass. More recently, \citet{Wetzel12} carried 
out a similar but more thorough analysis. They found that
the specific star formation rate (sSFR) of satellite galaxies 
in halos with different masses follows a bimodal distribution, but that  
the sSFR in the star forming sequence shows no strong 
dependence on halo mass.  Wetzel et al.  also investigated the 
excess in the quenched fraction of satellites relative to 
that of centrals, a quantity that reflects the strength of the impact 
of the intra-halo environments in quenching star formation, 
and found that this quantity increases with halo mass but
does not depend strongly on the stellar mass of the satellites.

On the theory side, galaxy formation models are still 
struggling to reproduce the basic observational results 
regarding the quenching of star formation by environmental effects. 
By comparing results obtained from galaxy groups with the 
prediction of the semi-analytical model (SAM) of \citet{Croton06}, 
\citet{Weinmann06b} suggested that the simple ``strangulation" 
model adopted by \citet{Croton06} and many other versions of SAM,
in which the diffuse gaseous halo is assumed to be stripped 
immediately after a galaxy is accreted into a larger halo, 
significantly over-quenches the satellite population.  
\citet{Font08} found that this ``over-quenching" problem
cannot be solved by a more physically motivated ``strangulation" 
model, but may be solved by assuming that the ejected gas from 
a satellite by supernova feedback is recycled back to the satellite 
to sustain further star formation. 

In a traditional SAM, a large number of prescriptions and model 
parameters are adopted to describe a variety of physical processes 
relevant to the formation and evolution of the galaxy population. 
This comprehensive nature of the model makes it hard to put tight 
constraints on specific physical processes, because of the 
degeneracy among different parts of the model
\citep[e.g.][]{Lu11a}. In particular, even the latest SAMs
still have trouble in reproducing the bimodal distributions
of the central galaxies in sSFR and color  and in their 
evolution \citep[e.g.][]{Lu14}. Since satellite galaxies
are believed to be evolved from centrals at higher redshift,    
the failure of a model in correctly reproducing the properties of central 
galaxies can also hinder the interpretation of the model prediction 
for the quenching of satellites. Several investigations have 
attempted to isolate the problem of satellite quenching  
from the total evolutionary process of galaxy evolution.  
For example, \citet{Balogh00} studied the halo-centric distribution 
of quenched satellites by assigning the SFRs of present-day 
centrals to satellites at the time of accretion and modeling 
the subsequent star formation of satellites with 
the Kennicutt law \citep[][]{Kennicutt98} and the cold gas 
reservoirs inherited from the progenitor centrals.  
\citet{Wetzel13} used a set of observational constraints 
to initialize the SFRs of satellites, and 
modeled the subsequent evolution of star formation in the 
satellites by assuming that a rapid quenching (an exponential decline
in SFR)  of a satellite occurs at a certain time after it is accreted by 
a larger halo. These phenomenological models can be used 
to translate observational constraints into certain physically 
meaningful quantities, such as the quenching time scale,  
but they themselves are not linked directly to any particular 
physical processes.

In this paper, we study the quenching of satellites by  
intra-halo environments. Our approach is different from 
those of previous investigations in two aspects.   
First, the stellar masses, star formation rates and gas contents
of the progenitors of satellite galaxies at the time of accretion 
are set by using the results obtained by \citet{LZ15b} from an  
empirical model constrained by a broad range of observations.
Second, the subsequent evolution of  satellite galaxies 
is modeled with physically motivated processes.
We test our model predictions by comparisons with results
obtained from groups of galaxies. The paper is arranged 
as follows. The properties of the progenitors of satellite
galaxies are described in  \S\ref{sec_progenitors}.
Our models for the evolution of satellite galaxies in 
dark matter halos are described in \S\ref{sec_satevolution}. 
Model predictions are presented in \S\ref{sec_prediction}, 
and a summary of our main results is given in  
\S\ref{sec_summary}. 

Throughout the paper, we use a $\Lambda$CDM cosmology with
$\Omega_{\rm m,0}=0.273$, $\Omega_{\Lambda,0}=0.727$, $\Omega_{\rm
b,0}=0.0455$, $h=0.704$, $n=0.967$ and $\sigma_{8}=0.811$.  
This set of parameters is from the seven year WMAP observations 
\citep{Komatsu11}. In addition, we adopt a \citet[][]{Chabrier03} IMF.


\section{Progenitors of Satellite Galaxies}
\label{sec_progenitors}

We start to trace the evolution of a satellite at the 
time when it was first accreted into a bigger halo. 
The stellar mass and star formation rate (SFR) 
are initialized according to the empirical model of 
\citet{LZ14, LZ15a} at the time of accretion. The empirical model 
links the galaxy stellar mass and SFR to dark matter 
halos of different masses  at different redshifts in a self-consistent way, 
and is constrained by the observed stellar mass functions 
measured from various redshift surveys.

Following convention, we refer a halo that has accreted into a bigger 
host halo as a sub-halo. The distribution of sub-halos in accretion redshift 
$z_{\rm acc}$ and in initial mass $M_{\rm s,i}$ is obtained from  
halo merger trees generated with the algorithm of \citet{Parkinson08}. 
This algorithm is based on the Extended Press-Schechter (EPS) formalism 
and tuned to match the conditional mass functions
of halo merger trees \citep{Cole08} constructed from the Millennium 
Simulation \citep[MS,][]{Springel05}.  As shown by 
\citet{Yang11} and \citet{Jiang14}, 
this algorithm is in good agreement with simulations in many other 
halo properties, such as mass assembly history, merger rate and 
un-evolved sub-halo mass function. For our analysis, merger trees 
are constructed over the redshift range $0\le z\le 15$,
with $100$ snapshots evenly distributed in $\ln(1+z)$ space.
The mass resolution is $2\times10^9\Msunh$.

Each sub-halo at the time of accretion is assigned a stellar mass 
and a star SFR according to its mass and accretion redshift
based on the results of \citet{LZ15a}. A cold gas mass is derived 
from its star formation rate using a star formation law, as described 
in \citet{LZ15b}.  In the present model, stars and cold gas are assumed to 
be distributed in thin disks. For simplicity, we assume that the surface 
densities of both the stellar and gas disks follow truncated 
exponential profiles. For the gas, it is
\begin{equation}
 \Sigma_{\rm g}\left(r\right) = 
  \begin{cases}
   \Sigma_{\rm g,0} \exp\left(-\frac{r}{R_{\rm g}}\right) & \text{if}\,\,\, r < R_{\rm tr,g} \\
   0 & \text{otherwise}
  \end{cases}
\end{equation}
where $\Sigma_{\rm g,0}$ is the central surface density,
$R_{\rm g}$ the scale radius, and $R_{\rm tr,g}$ the 
truncation radius. At the time of accretion of a satellite, 
the truncation radius is set to $\infty$. The subsequent evolution
of  the truncation radius is determined by different stripping 
processes, as described in the following sections.
The total mass within the truncation radius is 
\begin{equation}
 M_{\rm g} = 2 \pi R_{\rm g}^2 \Sigma_{\rm g,0}
             \left[1 - \left(1+\frac{R_{\rm tr,g}}{R_{\rm g}}\right) 
                   \exp\left(-\frac{R_{\rm tr,g}}{R_{\rm g}} \right) 
             \right]\,.
\end{equation}
The stellar disks are assumed to follow a profile similar to the 
gas but with a different scale radius $R_{\star}$ and 
a different truncation radius $R_{\rm tr,\star}$.
Both $R_{\star}$ and $R_{\rm g}$ are set using the model described in 
\citet{LZ15b}, in which the ratio $\mathcal{L} \equiv R_{\rm g}/R_{\star}$ 
is assumed to be a constant larger than one.

The initial gas mass $M_{\rm g, i}$ is obtained from the initial star formation rate
(SFR) through an adopted star formation law. Specifically, we write 
\begin{equation}
 {\rm SFR} = \int_{0}^{\infty} \Sigma_{\rm SF}\left(\Sigma_{\rm g}\right)2\pi R{\rm d}R\,,
\end{equation}
where $\Sigma_{\rm SF}$ is the SFR surface density, assumed to be 
determined by the cold gas mass surface density, $\Sigma_{\rm g}$. 
In this paper we implement the Kennicutt-Schmidt 
law \citep{Kennicutt98},\footnote{We have made a test 
using the star formation model of  
\citet{Krumholz09}. The results are qualitatively the same.} which gives
\begin{equation}
 \label{kennicutt_a}
 \Sigma_{\rm SFR} = 
 \begin{cases}
  A_{\rm K} \left(\frac{\Sigma_{\rm g}}{\rm M_{\odot}pc^{-2}}\right)^{N_{\rm K}}\,\,\,\,\text{if}\,\,\, \Sigma_{g} \ge \Sigma_{\rm c},\\
  0\,\,\,\,\text{if}\,\,\, \Sigma_{\rm g} < \Sigma_{\rm c}\,.
 \end{cases}
\end{equation}
The power index $N_{\rm K} \approx 1.4$, and the amplitude
$A_{\rm K} \approx 2.5\times10^{-4}\Msun {\rm yr^{-1}pc^{-2}}$.
These parameters are constrained using the gas mass - stellar mass relation
of local galaxies \citep[and references therein]{LZ15b}. 

\section{Evolution of Satellite Galaxies}
\label{sec_satevolution}

\subsection{Orbits of Satellites}
\label{ssec_orbits}

The orbits of satellite galaxies in host dark matter halos 
are modeled according to the orbits of sub-halos within which the 
satellites are located.  The host halos are assumed to follow the 
NFW profile \citep{Navarro96}:
\begin{equation}
 \rho_{\rm D}(r) = \frac{M_{\rm vir}}{4\pi R_{\rm vir}^3}
                \frac{c^2}{\ln\left(1+c\right) - c/\left(1+c\right)}
                \frac{1}{x\left(1+cx\right)^2}
\end{equation}
where $M_{\rm vir}$ is the halo mass, $R_{\rm vir}$ the  
radius of the halo, and $x \equiv r/R_{\rm vir}$. 
The concentration parameter,  $c$, is modeled with the 
$c$-$M_{\rm vir}$ relation given by \citet{Bullock01}.
Within the mass and redshift ranges we are interested in,
the model is a reasonably good approximation to the more 
recent models of the halo concentration - mass relation  
\citep[e.g.][]{Zhao09, Dutton14}. The corresponding gravitational 
potential of this density profile is 
\begin{equation}
 \Phi_{\rm D} (r)= -V_{\rm vir}^2 
        \frac{1}{\ln\left(1+c\right) - c/\left(1+c\right)}
        \frac{\ln\left(1+cx\right)}{x},
\end{equation}
where $V_{\rm vir}^2\equiv GM_{\rm vir} /R_{\rm vir}$ is 
the virial speed of the halo. 

At the time of accretion, that is when a sub-halo crosses the 
virial radius ($R_{\rm vir}$) of a larger halo, 
the infall velocity of the sub-halo is set to be exactly the virial 
velocity $V_{\rm vir}$. This is a natural prediction
of the spherical collapse model, and is used here as an 
approximation. The initial orbital angular momentum 
of the sub-halo is described by its orbital circularity, $\eta$, 
which is defined to be the ratio between the true orbital angular 
momentum of the sub-halo and the orbital angular momentum of a circular 
orbit with the same orbital energy as the sub-halo. 
The distribution of this quantity has 
been determined using high resolution simulations. Here we use 
the result of \citet{Zentner05}, which gives 
\begin{equation}
P(\eta)\propto \eta^{1.2}(1-\eta)^{1.2}\,.
\end{equation}

The subsequent evolution of the sub-halo is traced by following
the model of \citet{Taylor01}, which includes dynamic friction and 
tidal stripping.  As the sub-halo moves through the 
background of dark matter particles, a drag force (dynamic friction)
acts to reduce its angular momentum and energy. 
This drag force is modeled by Chandrasekhar's formula,
\begin{eqnarray}
 \label{df}
 {\bf F}_{df} & = & -4\pi G^2 M_{\rm s}^2 \ln\Lambda \rho_{\rm D}( {\bf r}) \\ \nonumber
              &   & 
                \left[
                 {\rm erf}\left( \frac{v_{\rm s}}{V_{\rm c}} \right)
                 - \frac{2}{\sqrt{\pi}} \frac{v_{\rm s}}{V_{\rm c}} 
                  {\rm e}^{-(v_{\rm s}/V_{\rm c})^2}\right]
                \frac{{\bf v_{\rm s}}}{v_{\rm s}^3}\,,
\end{eqnarray}
where $M_s$ is the instantaneous mass of the sub-halo, 
${\bf v}_{\rm s}$ is its velocity, and $V_{\rm c}$
is the local circular velocity of the host halo. The Coulomb 
logarithm is estimated as $\ln\Lambda= \ln(M_{\rm vir}/M_{\rm s})$,
where $M_{\rm vir}$ is the mass of the host halo.

The instantaneous tidal radius of the sub-halo is estimated using 
\begin{equation}
 R_{\rm tidal} \approx 
      \left(
        \frac {GM_{\rm s}\left(<R_{\rm tidal}\right)} {\omega^2-\frac{\partial^2 \Phi\left(r\right)}{\partial r^2}}
      \right)^{1/3},
\end{equation}
where $\omega$ is the instantaneous orbital angular velocity. This tidal radius 
is calculated at each time step as the orbit of the satellite is integrated. 
Following \citet{Taylor01}, the mass outside the tidal radius is assumed to be stripped
at the rate given by
\begin{equation}
 \dot{M_{\rm s}} = \frac{\omega}{2\pi} M_{\rm s}\left(\ge R_{\rm tidal}\right)\,.
\end{equation}
The sub-halo mass so obtained is then used in the dynamic friction 
calculation at each time step. Note that the disk contained in a sub-halo
is also assumed to be tidally truncated, so that the disk truncation 
radius, $R_{\rm tr,g}$, introduced in \S\ref{sec_progenitors}, is at 
most $R_{\rm tidal}$. 

The orbit of each sub-halo (and of the galaxies it contains) 
is integrated starting from the time of 
accretion with the initial conditions specified above, taking 
into account both gravity and dynamical friction. 
We stop tracing the orbit of a sub-halo when it has lost
all its orbital angular momentum. At this point, 
the satellite galaxy associated with the sub-halo is assumed 
to  have merged with the central galaxy of the host halo.

\subsection{High-order Substructures}

When a group of galaxies merges with a more massive system,
some of the galaxies and substructures may still be associated 
with their parent halos, becoming high order substructures. 
Others may be tidally stripped, moving on independent orbits
in the new, bigger halo. Whether a substructure will become a high-order  
substructure after merging depends on how they are gravitationally 
bound to each other as well as on the orbit of their parent halo.
An accurate treatment of the high-order substructures, 
therefore, requires a recursive scheme: we need to trace not only the 
trajectories of the sub-halos but also the trajectories of the
sub-sub-halos within them. A detailed treatment of this makes the 
model very complicated and is beyond the goal of this paper. 
Instead,  we adopt the pruning scheme developed by 
\citet{Taylor04}, in which the substructures that are loosely bound
to the parent halo are assumed to be directly stripped when the 
merger happens.  An empirical criterion suggested by the authors
is that if a substructure has spent less than a period of $n_{\rm 0}P_{\rm rad}$
in the parent halo, it is considered loosely bound, where 
$n_{\rm 0} = 2$ is a free parameter calibrated using more 
accurate models, and $P_{\rm rad}$ is the mean orbital period.
Each of the stripped sub-halos is assigned a new orbital parameter 
and is followed in the same way as the other sub-halos in the new halo.  
About $70\%$ of the substructures are stripped from the parent 
halos according to this model. The remaining, tightly bound part 
is assumed to stay with their parent until the parent itself is 
destroyed.

\subsection{Star Formation and Outflow}

After a satellite is accreted by a larger halo, its cold gas can be 
depleted by different processes. The first is star formation and 
galactic outflows driven by the star formation. 
The gas mass depletion rate due to this process is
\begin{equation}
 \dot{M}_{\rm g} = \left(1-\mathcal{R}+\epsilon_{\rm w} \right)
   \int_{0}^{R_{\rm tr,g}} \Sigma_{\rm SF}\left(\Sigma_{\rm g}\right)2\pi R{\rm d}R\,,
\end{equation}
where $\Sigma_{\rm SF}\left(\Sigma_{\rm g}\right)$ is 
the star formation law, $\mathcal{R}$ is the recycled fraction
from stellar evolution and $\epsilon_{\rm w}$ is the loading factor
of the mass outflow driven by star formation.
For simplicity, we assume that this process 
changes neither the scale radius nor the truncation radius
of the disk, but only reduces the central surface density $\Sigma_{\rm g,0}$, 
so that 
\begin{equation}
 \frac{\dot{\Sigma}_{\rm g,0}}{\Sigma_{\rm g,0}}
  = \frac{\left(1-R+\epsilon_{\rm w}\right) SFR}{M_{\rm g}}\,.
\end{equation}
The star formation in the gas disk is assumed to follow the same 
model as described in \S\ref{sec_progenitors}.

In addition to star formation and outflow, other processes,  
such as ram-pressure stripping and tidal interactions, may also deplete 
the disk gas outside a certain radius at which the
external forces are balanced by the gravitational potential of the 
satellite itself. We assume that the stripping does not change the inner 
structure of the disk, so that $\Sigma_{\rm g,0}$ and $R_{\rm g}$
are not affected. The detailed modeling of these processes are 
described in the following.

\subsection{Ram Pressure Stripping}
\label{sec_rampressure}

The host halos are assumed to be filled with diffuse hot gas.
The total mass of the diffuse gas within the virial radius is modeled 
on the basis of the hot mode accretion presented in \citet{Keres05} and 
\citet{Lu11}, and the hot gas fraction is given by 
\begin{equation}
 f_{\rm hot} = \frac{1}{2} 
              \left(
               1+{\rm erf}\left[
                           \frac{\log_{10}\left(M_{\rm h}/M_{\rm tran}\right)} {\sigma_{\rm tran}}
                          \right]
              \right)
\end{equation}
where $\log_{10}\left(M_{\rm tran}\right) = 11.4$ is the transition halo
mass, and $\sigma_{\rm tran} = 0.4$.

The diffuse gas is assumed to be isothermal and has the 
virial temperature $\frac{1}{2} (\mu m_{\rm p}/ k_{\rm b}) V_{\rm vir}^2$, 
where $\mu$ is the mean molecular weight of the gas. 
Assuming hydrostatic equilibrium, the density profile of the 
diffuse gas is determined by the gravitational potential of the
host halo,
\begin{equation}
 \rho_{\rm X}(r)  = \rho_{\rm X,0} 
   \exp\left[
    -\frac{\mu m_{\rm p}}{k_{\rm B}T_{\rm vir}} \Phi_{\rm D}\left(r\right)
   \right]
\end{equation}
The integral constant $\rho_{\rm X,0}$ is given by the total mass 
of the diffuse gas through 
\begin{equation}
f_{\rm hot} \frac{\Omega_{\rm b,0}}{\Omega_{\rm m,0}} 
M_{\rm vir} =\int_{0}^{R_{\rm vir}} \rho_{\rm X}(r) 4\pi r^2{\rm d}r\,.
\end{equation}

The gravitational restoring force per unit area of a infinitesimally thin disk is
\begin{equation}
 P_{\rm gr} = 2 \pi G \Sigma_{\star} \Sigma_{\rm g},
\end{equation}
where $\Sigma_{\star}$ and $\Sigma_{\rm g}$ are the surface 
densities of stars and gas at the radius in question, respectively.
For a face-on disk, the truncation radius owing to ram pressure stripping
can be calculated by setting the gravitational restoring force to be
equal to the ram pressure $\rho_{\rm X}\left(r\right)v_{\rm s}^2$,
which gives
\begin{equation}
 \label{ram_pressure_0}
 R_{\rm rp} = \frac{R_{\star}R_{\rm g}}{R_{\star}+R_{\rm g}} 
              \ln \left(
              \frac{2 \pi G \Sigma_{\star,0}\Sigma_{\rm g,0}}
              {\rho_{\rm X}\left(r\right) v_{\rm s}^2}
              \right)\,.
\end{equation}
In general, if the disk is inclined relative the 
moving direction, the ram pressure exerted on the disk depends
on the inclination angle.  Analytically deriving this inclination dependence 
is difficult, as the stripping becomes asymmetric when the 
disk in inclined. Using a set of hydrodynamic simulations, \citet{Roediger06} 
measured the mean truncation radius as a function of 
both the ram pressure and the inclination angle. It is  
found that the prediction of Eq.\,(\ref{ram_pressure_0}) only deviates
from the simulations when the inclination angle 
$\theta \ge 60^{\circ}$. To take this into account, we add 
a correction term to Eq.\,(\ref{ram_pressure_0}) so that it is 
consistent with the simulation result:
\begin{equation}
 \label{ram_pressure_1}
 R_{\rm rp} = \frac{R_{\star}R_{\rm g}}{R_{\star}+R_{\rm g}} \left[
              \ln \left(
              \frac{2 \pi G \Sigma_{\star,0}\Sigma_{\rm g,0}}
              {\rho_{\rm X}\left(r\right) v_{\rm s}^2}
              \right)
              -0.8\ln\left(\cos \theta\right) \right]\,,
\end{equation}
where $\cos\theta = \vert{\bf v_{\rm s}}\cdot{\bf s}\vert/v_{\rm s}$.
The spin axis ${\bf s}$ of the disk is assigned at the time 
of accretion and is assumed to be independent of any other 
quantities. 

In addition to ram pressure stripping, gas disk can also
lose mass because of Kelvin-Helmholtz instability. However, 
according to \citet{Roediger07}, this mass loss rate is far less 
efficient than ram pressure stripping, and so it will be ignored
in our model.

\subsection{Tidally Triggered Starburst}

Interaction between galaxies is another possible process that can 
lead to the quenching of star formation. As a satellite galaxy passes by the
central galaxy, the tidal field from the central region can strip
the outskirts of the disk of the satellite. This direct tidal stripping 
is modeled with the tidal radius described in \S\ref{ssec_orbits}.  
In addition, the tidal interaction can also exert
a torque on the disk, causing the disk gas to lose angular 
momentum and funneling it into the nuclear region 
to trigger a star burst.  This process may also quench the 
satellite by consuming the star forming gas quickly. 

The enhancement of the star formation in flyby satellites
has been found both observationally \citep{Li08} and 
theoretically \citep{DiMatteo07}. In particular, \citet{Li08} found
that, for star-forming galaxies, 
(i) the enhancement of star formation in the flyby satellites
occurs when the projected distance is less than $5r_{\rm 90}$,
where $r_{\rm 90}$ is the radius that encloses $90\%$ of the light of 
the central galaxies;  (ii) the enhancement is a factor of about $3$;  
(iii) the results depend only weakly on the mass of the companion. 

Motivated by the findings of \citet{Li08},  we consider a simple 
phenomenological model to emulate the star burst triggered by 
tidal interaction. We assume that the tidal interaction becomes 
important only when a satellite is within some radius from the 
halo center. This radius is assumed to scale with the size of 
the central galaxy as $R_{\rm sb} = x_{\rm sb} R_{\star}$,
where $R_{\star}$ is the scale radius of the stellar disk  and is 
estimated following the redshift-dependent galaxy size - stellar 
mass relation as described in \citet{LZ15b}. 
When a satellite reaches $R_{\rm sb}$, it is 
switched to a ``star burst'' mode, in which the SFR
is assumed to be enhanced by a factor of    
$\mathcal{E}_{\rm enhance}$ relative to its SFR immediately before 
it reaches $R_{\rm sb}$. This enhanced SFR is assumed 
to last until the gas is completely exhausted or the satellite gets
out of the sphere, whichever comes first. If the gas is not
exhausted before the satellite moves out of the sphere, 
the galaxy will resume its pre-burst mode of star formation, 
but with SFR calculated from the reduced amount of cold gas.  
We also assume that the star burst generates an 
outflow with a loading factor $\epsilon_{\rm w, sb}$, which 
may be different from $\epsilon_{\rm w}$.

\subsection{Environmental Quenching versus Self-Quenching}

Quenched galaxies exist in a variety of environments, 
although satellites tend to have a higher fraction of 
quenched population than centrals of similar stellar masses. 
It is believed that the physical 
processes responsible for the quenching of centrals are not 
related to environments. A widely adopted 
scenario is quasar-mode feedback, which can
eject the cold gas in a galaxy and quench its star formation in a very 
short time scale. Such mechanisms are usually referred
to as ``self-quenching'', to distinguish them from environmental 
processes. Self-quenching is expected to have also operated 
in the observed population of satellites before or after 
their accretions by larger halos. If one assumes that 
self-quenching and environmental quenching are 
independent of each other,  the excess of quenched 
fraction in satellites relative to that in centrals can be used as an 
indicator of the efficiency of environmental quenching. 
This excess can be described by the ratio 
\begin{equation}
  f_{\rm Q, excess}^{\rm sat} = 
  \frac{f_{\rm Q,now}^{\rm sat}-f_{\rm Q,now}^{\rm cen}}
  {f_{\rm A,now}^{\rm cen}}\,,
\end{equation}
where $f_{\rm Q}$ denotes the quenched fraction,
$f_{\rm A}$ the total  fraction, and the superscripts, `cen' and 
`sat', indicate centrals and satellites, respectively. This quantity
has been estimated by \citet{Wetzel12, Wetzel13} using 
galaxy groups selected from the SDSS. In our investigation, we will
focus on environmental quenching without modeling the mechanisms 
that are responsible for the quenching of centrals.  In what follows, 
we will refer  $f_{\rm Q, excess}^{\rm sat}$ as the quenched 
fraction without including the population that is quenched
before the galaxies become satellites.   
\begin{figure*}
 \centering
 \includegraphics[width=0.5\linewidth]{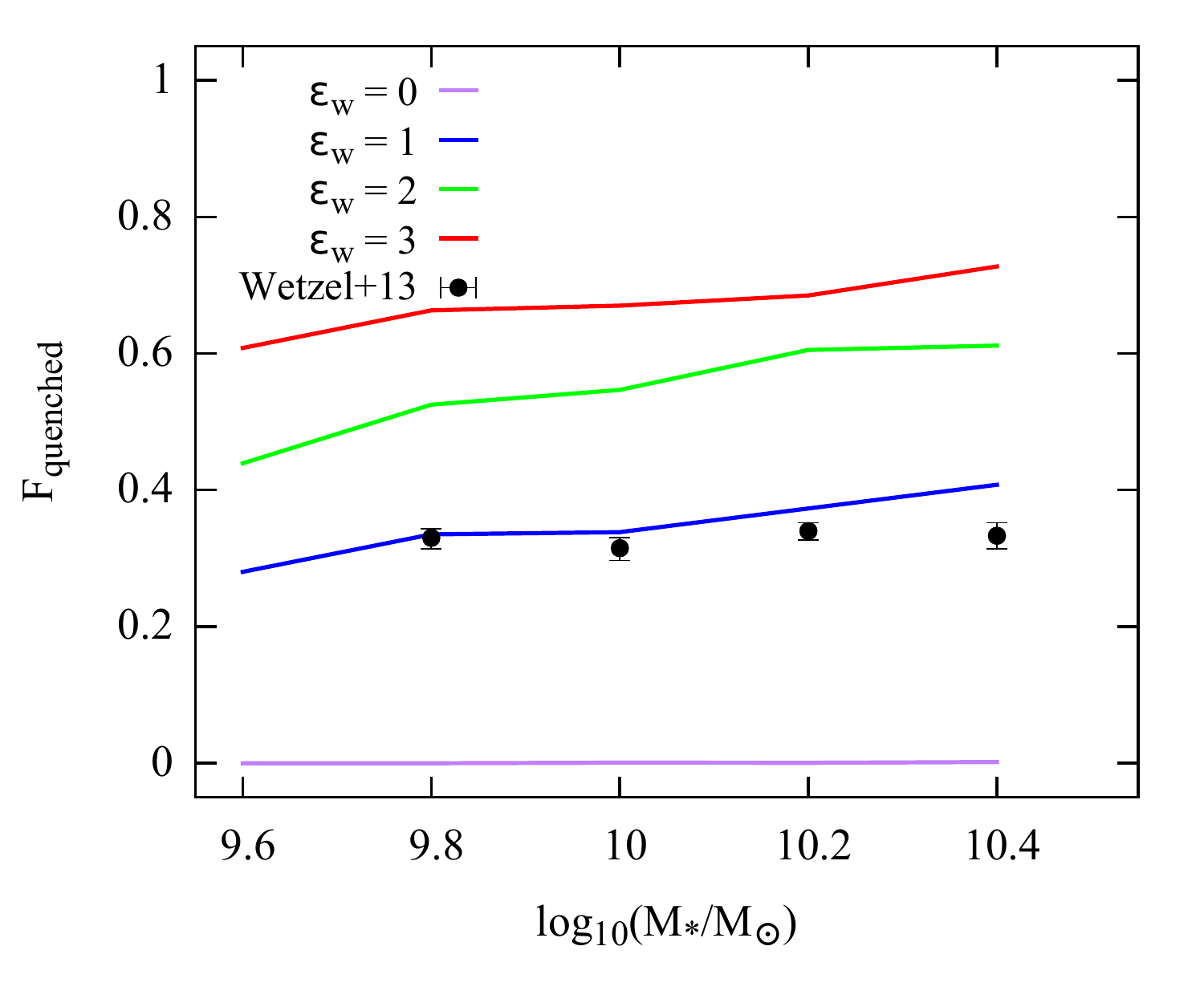}
 \caption{The colored lines are the fractions of quenched galaxies
          predicted by the model 
          in which disk gas is consumed 
          only by star formation and associated outflow. 
          Different colors represent different outflow 
          mass loading factors,
          as indicated in the panel.
          The data points are taken from \citet{Wetzel13}.}
 \label{Fq_Msat_RPoff}
\end{figure*}

%
\section{Model Predictions}
\label{sec_prediction}

\subsection{Quenching by Strangulation}

In the literature, strangulation refers to a hypothesis that 
a galaxy in a sub-halo will stop accreting new gas as soon as 
the sub-halo is accreted into a larger halo.  Because of strangulation, 
star formation in satellite galaxies will gradually slow down 
and eventually stop as the gas reservoir is consumed by 
star formation, even in the absence of any other additional 
processes. Galactic wind driven by star formation speeds up 
the depletion of the cold gas. This is generally how red satellites 
are produced in many semi-analytical models (SAMs) in the 
literature \citep[e.g.][]{Lu14}. To understand how this process
works, we consider a model in which all other environmental 
processes (ram pressure stripping and tidally triggered 
starburst) are switched off, so that we can focus on quenching 
by gradual gas consumption owing to star formation and 
the associated outflow.   

\begin{figure*}
 \centering
 \includegraphics[width=0.9\linewidth]{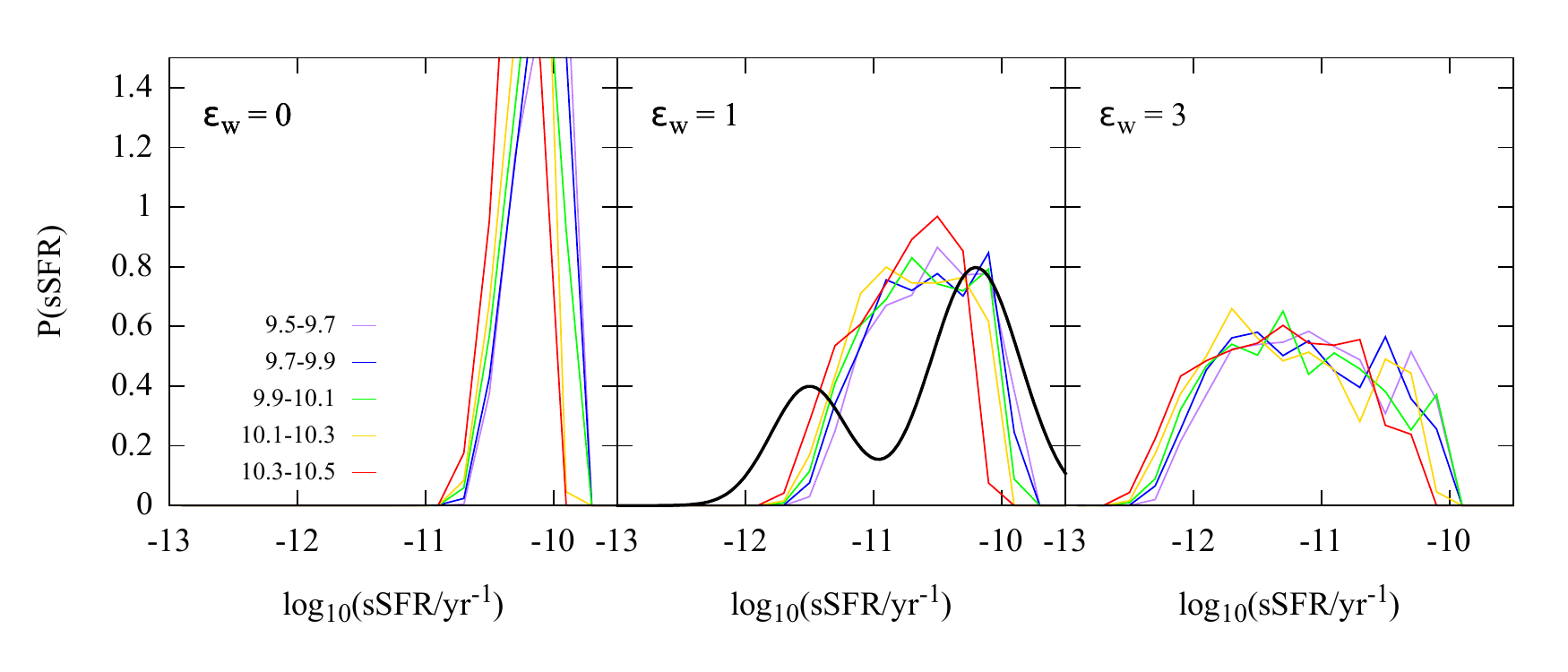}
 \caption{The distribution of the sSFR of satellite galaxies
          predicted by the model
          in which disk gas is consumed 
          only by star formation and associated outflow. 
          Different panels correspond to different outflow mass loading factors.
          The stellar mass range of satellites is coded in different colors, as
          indicated in the left panel. The black solid line in the middle
          panel is the observed distribution inferred from \citet{Wetzel12,Wetzel13}.}
 \label{sSFR_KS_RPoff}
\end{figure*}

Figure\,\ref{Fq_Msat_RPoff} shows the impact of the outflow loading
factor, $\epsilon_{\rm w}$, on the predicted quenched fraction of satellites.
All the satellites stay as star forming galaxies if $\epsilon_{\rm w} $ is 
set to be 0. This means that the cold gas reservoir left over from 
the time of accretion is sufficient to keep satellite galaxies active 
in star formation all the way to the present day if there is no 
other gas depletions than star formation. The observational 
constraints seem to favor $\epsilon_{\rm w} = 1$ 
and adopting a loading factor $\epsilon_{\rm w} > 1$ 
tends to over-quench the satellite population.
Unfortunately, the loading factor of outflow is still poorly understood.
In hydrodynamic simulations and SAMs, a large value seems 
to be required to lower the efficiency of star formation to reconcile 
the tension between the observed galaxy luminosity/stellar mass function
(dominated by central galaxies) and the halo mass function
at the low-mass end \cite[e.g.][]{Yang03}.
For instance, in the Somerville model
studied in \citet{Lu14}, the loading factor 
is about $3$ for a Milky-Way like halo and 
is more than $10$ for a 
$10^{11}\Msun$ halo. In many SAMs, the outflow loading factor
assumed for satellite galaxies is the same (high value) as for 
centrals, and so the satellite population is expected to be 
over-quenched according to the results shown in     
Figure\,\ref{Fq_Msat_RPoff}.

Indeed, satellite galaxies are predicted to be too red, 
with too little current star formation, in many SAMs
\citep[e.g.][]{Lu14},  and strangulation is usually blamed for this
over-quenching problem \citep{Weinmann06b}. 
However, detailed modeling of the stripping of the diffuse gas halo
suggests that the problem cannot be solved by continuous accretion 
of fresh gas from the parent halo \citep{Weinmann10, Font08}.
In order to solve this problem,  \citet{Kang08} 
and \citet{Weinmann10} 
had to assume that ram pressure does not make any 
significant stripping of the diffuse halo gas.  Indeed, \citet{Font08}  
assumed that most of the gas reheated by star formation - driven 
wind in a satellite will not be stripped, but will instead be re-accreted by 
the satellite, so that the gas reservoir originally associated with the satellite 
is always available for star formation, even though large amounts 
of gas is assumed to be in the wind.  

The assumption of strong galactic winds for satellite galaxies 
needs to be revisited, however. In fact, to match theory with 
observation does not require a large loading factor for central 
galaxies. For instance, using empirically constrained star formation-halo 
mass relation and the gas phase metallicity, \citet{LZ15b} found  
that the mass loading factor of galaxy wind required to suppress 
star formation in present-day central galaxies is no larger than unity if 
the accretion of pristine gas into a halo can be suppressed 
by preheating.  If satellite galaxies inherit the same star 
formation and feedback from their progenitors,  then there 
will be no over-quenching problem, as shown above. 

As shown in Figure\,\ref{sSFR_KS_RPoff}, the specific 
star formation rate (sSFR, defined to be the ratio between star 
formation rate and stellar of of a galaxy) distribution predicted by this 
gradual gas depletion model is always unimodal, and increasing 
the loading factor only broadens the distribution without changing 
the unimodal nature.  This is in contrast with the observational 
results that the sSFR of satellite galaxies follows a bimodal 
distribution (shown by the smooth thick solid curve in the middle 
panel) and such a distribution persists in different environments 
\citep[e.g.][]{Wetzel13}. This demonstrates that strangulation
alone cannot explain the observed quenching of satellite galaxies.
\begin{figure}
 \centering
 \includegraphics[width=0.9\linewidth]{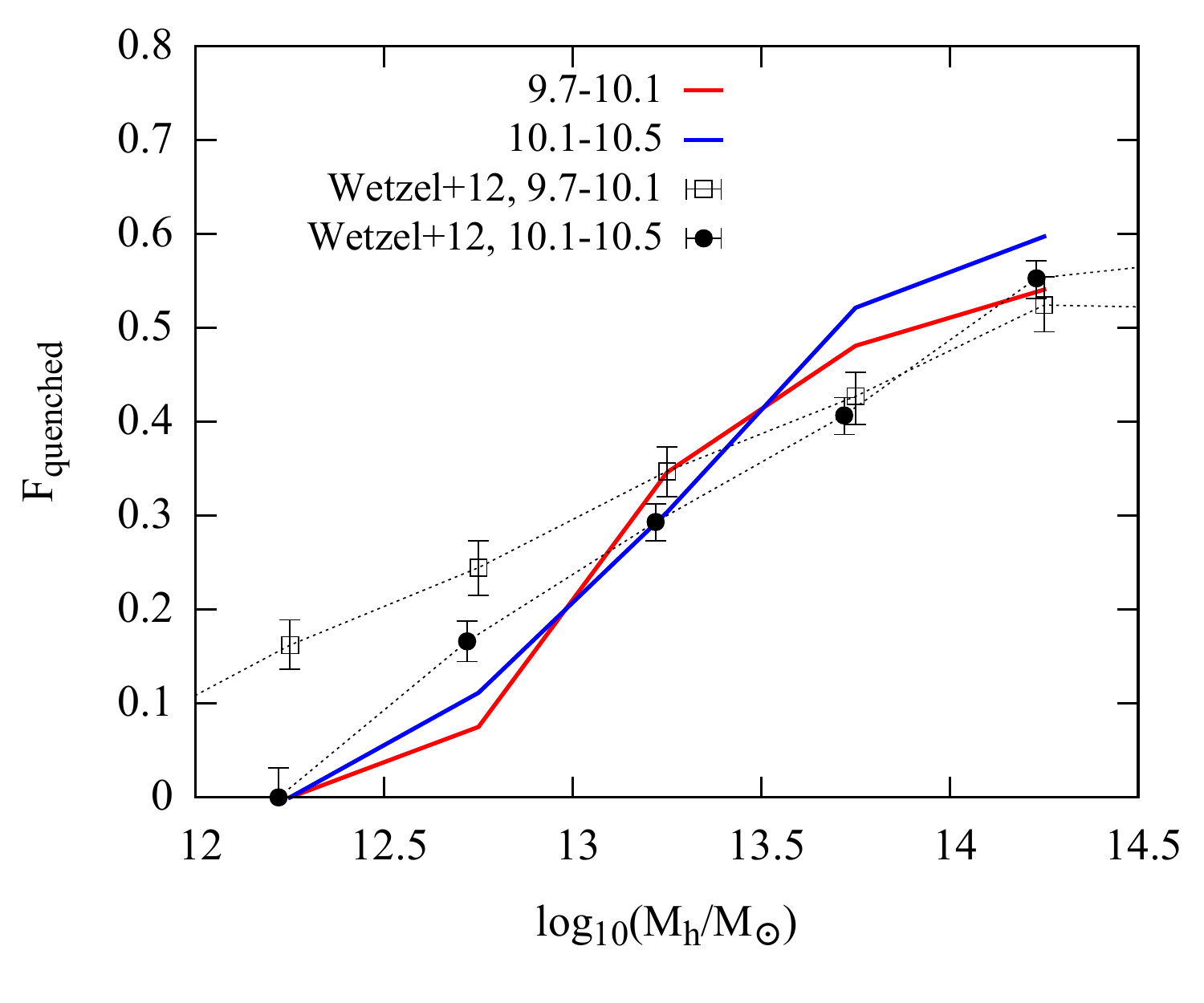}
 \caption{Fraction of quenched satellites as a function host halo mass, 
          predicted by the model
          in which disk gas is consumed 
          only by star formation and associated outflow. 
          The solid colored lines are predictions by the model 
          assuming $\epsilon_{\rm} = 1$.
          The data points with error bars are taken from \citet{Wetzel12}.}
 \label{Fq_Mh_RPoff}
\end{figure}

Figure\,\ref{Fq_Mh_RPoff} shows the quenched fraction 
as a function of host halo mass.  Here results are only shown 
for the $\epsilon_{\rm w}=1$ model, as it correctly reproduces
the overall quenched fraction. The quenched fraction increases 
with halo mass, because satellites of the same mass were 
accreted earlier in more massive halos and so a larger fraction of
them have exhausted their gas by present day. The model under-predicts 
the quenched fraction in halos with masses $<10^{13}\Msun$ in 
comparison with the observational data, demonstrating again that 
strangulation alone cannot explain the observed quenching of 
satellite galaxies in different halos, even though the outflow loading factor 
is tuned so that the model correctly predicts the overall quenched fraction 
of the satellite population.  

\begin{figure}
 \centering
 \includegraphics[width=0.9\linewidth]{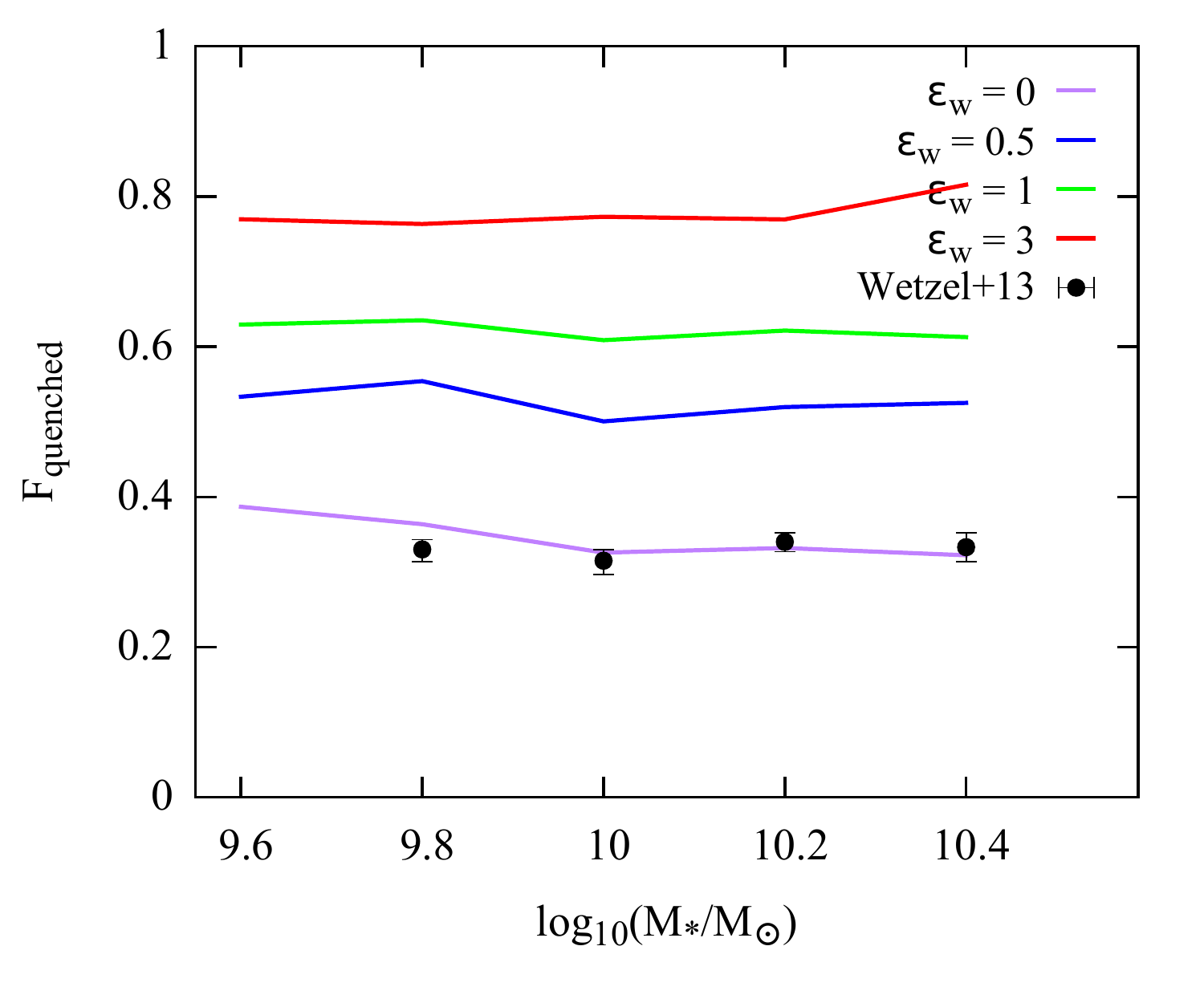}
  \caption{The predicted fractions of quenched galaxies
          by the model with
          ram-pressure stripping.
          Different colors represent different outflow 
          mass loading factors,
          as indicated in the panel.
          The data points are taken from \citet{Wetzel13}.}
 \label{Fq_Msat_RPon}
\end{figure}

\subsection{Quenching by Ram-Pressure Stripping}
\label{ssec_rampressure}


\begin{figure*}
 \centering
 \includegraphics[width=0.9\linewidth]{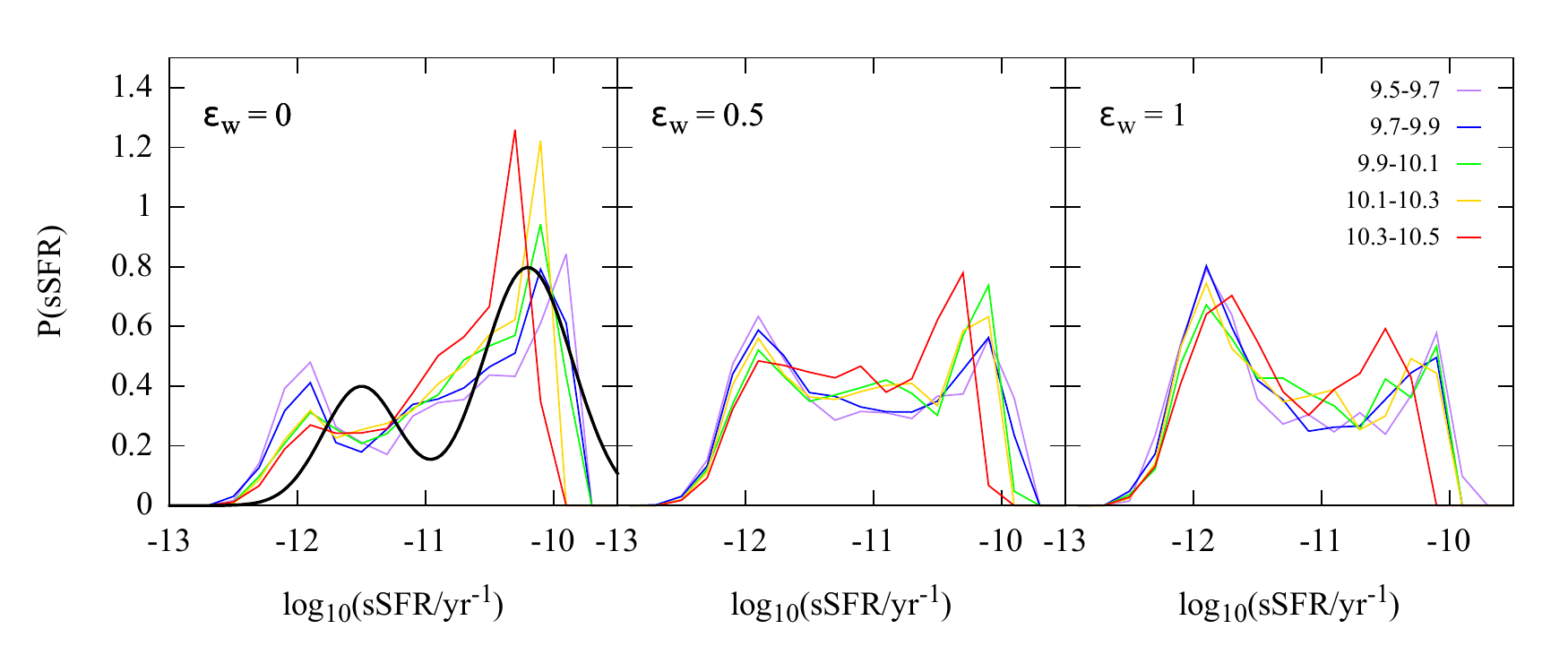}
 \caption{The distribution of sSFR of satellite galaxies
          predicted by the model of ram-pressure stripping.
          Different panels correspond to different assumed 
          outflow mass loading factors.
          The stellar mass range of satellites is coded in different colors, as
          indicated in the right panel. The black solid line in the first
          panel is the observational distribution inferred from \citet{Wetzel12,Wetzel13}.}
\label{sSFR_KS_RPon}
\end{figure*}
\begin{figure*}
 \centering
 \includegraphics[width=0.8\linewidth]{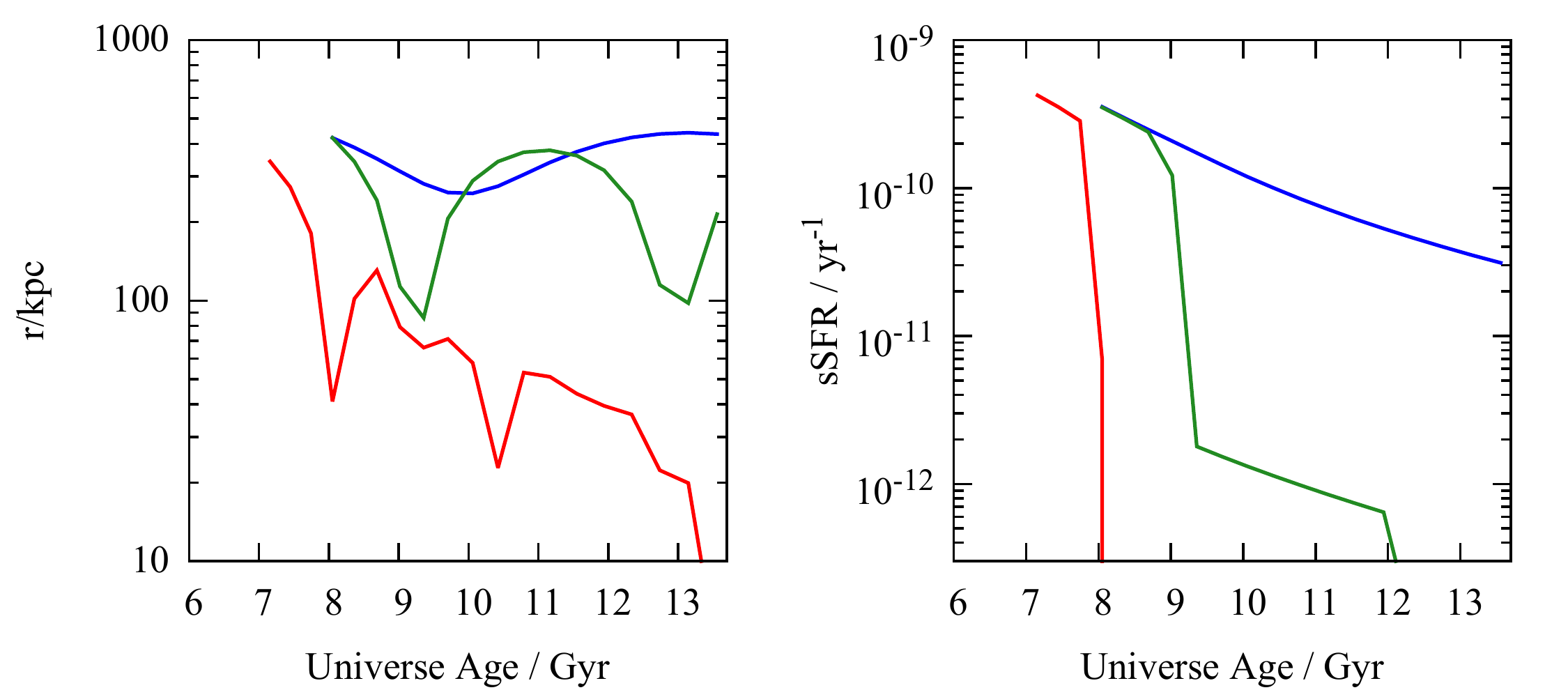}
 \caption{
  The evolution history of three satellites on three different 
  initial orbits (red: nearly radial; blue nearly circular; green:
  intermediate).  
  The left panel shows  the distance to the host center 
  as a function of time, and
  the right panel shows the specific star formation rate
  as a function of time. 
  The host halos are of $10^{13}\Msunh$ at present day.
 }
 \label{trajectory}
\end{figure*}
\begin{figure*}
 \centering
 \includegraphics[width=0.5\linewidth]{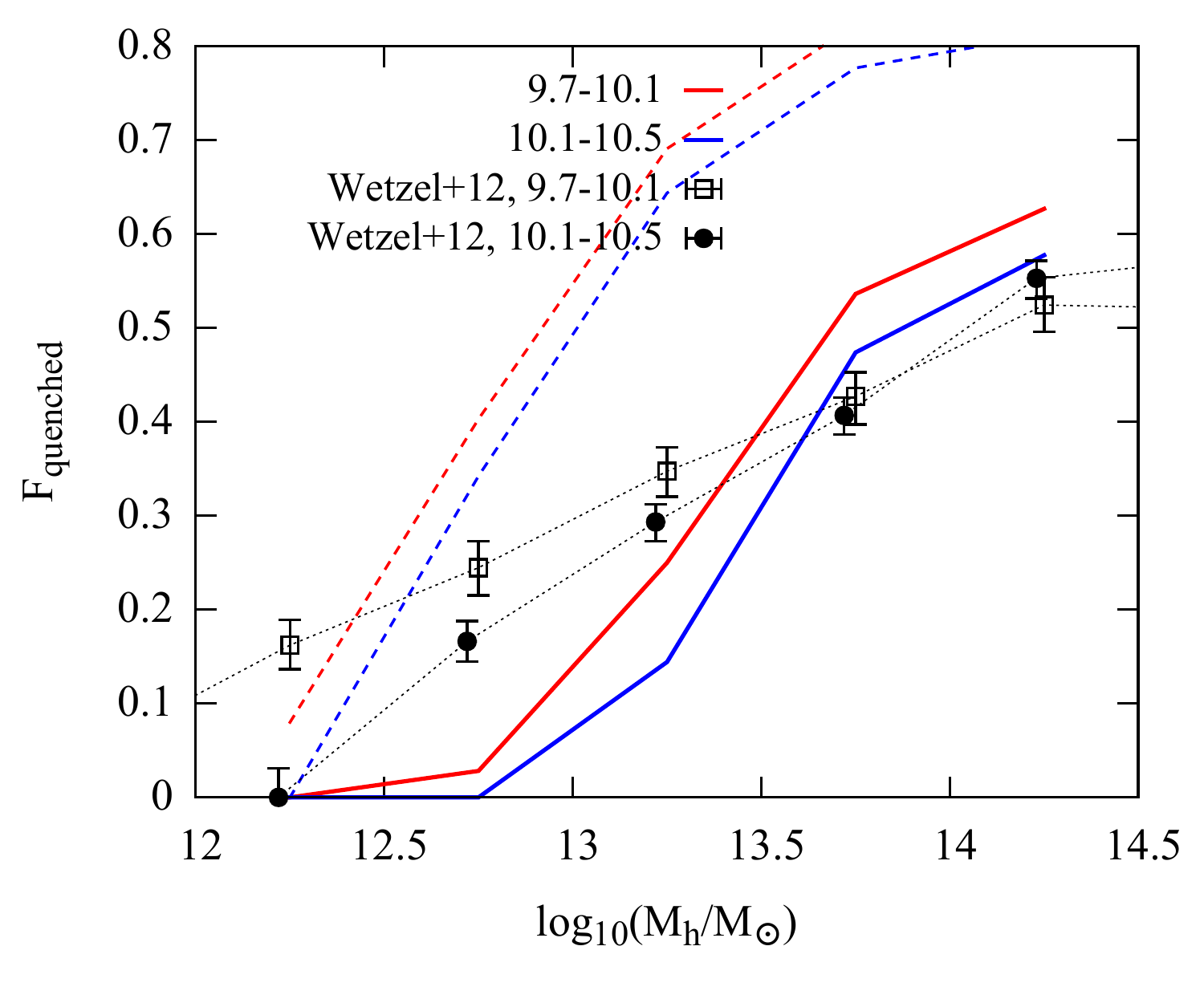}
 \caption{The fraction of quenched satellites as a function host halo mass.
          The colored solid lines are predictions by the model with ram pressure 
          stripping and with $\epsilon_{\rm} = 0$,
          and the colored dashed lines are predictions of the model
          with $\epsilon_{\rm} = 1$.
          Data points with error bars are taken from \citet{Wetzel12}.}
 \label{Fq_Mh_RPon}
\end{figure*}
\begin{figure*}
 \centering
 \includegraphics[width=0.5\linewidth]{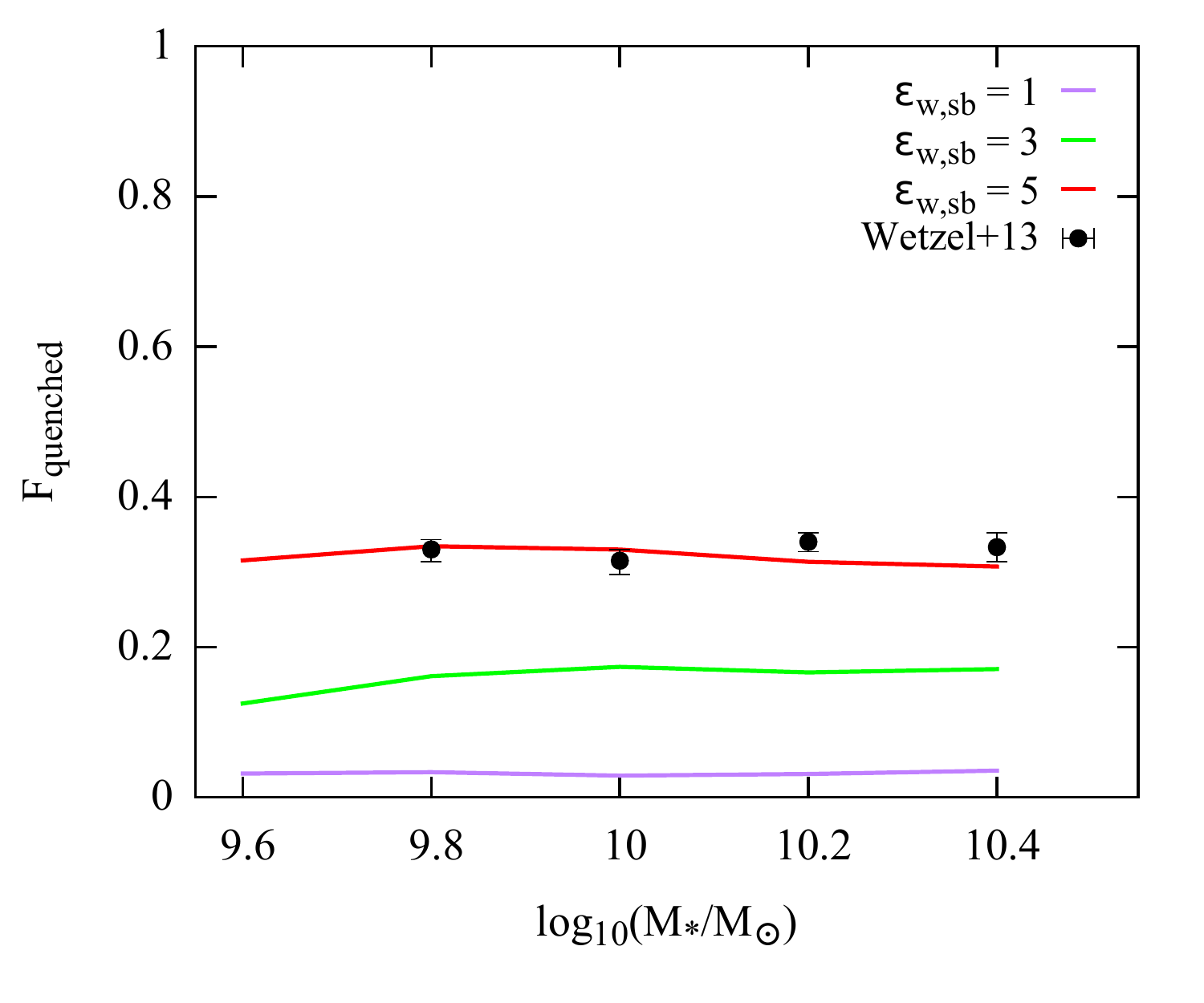}
   \caption{The predicted fractions of quenched galaxies
          by the model of tidally-triggered starburst. 
          Different colors represent different outflow 
          mass loading factors assumed for the starburst 
          phase, as indicated in the panel.
          The data points are taken from \citet{Wetzel13}.}
\label{Fq_Msat_SB}
\end{figure*}

Next we include ram-pressure stripping in a way described in 
\S\ref{sec_rampressure}. Figure\,\ref{Fq_Msat_RPon} shows 
the overall quenched fraction predicted by this model and 
how it  changes with the assumed outflow loading factor. As one can see, 
when ram-pressure stripping is included, any loading factor 
$\epsilon_{\rm w} > 0$ tends to over-quench the satellite population, 
and the model prefers $\epsilon_{\rm w} =0$, i.e. no outflow. 
Setting $\epsilon_{\rm w}$ to $1$,  the model already over-redicts 
the quenched fraction by $80\%$.
The two depletion mechanisms, outflow and ram-pressure stripping,  
do not act independently. Galactic wind depletes the gas at the 
central region of the disk, where most stars form. This reduces the 
gas surface density, thereby reducing the gravitational restoring 
force and making the ram pressure stripping more effective
[see Eq.\,(\ref{ram_pressure_1})].  

Unlike gradual gas consumption, ram pressure stripping 
produces a bimodal distribution in the sSFR, as shown 
in Fig.\,\ref{sSFR_KS_RPon}. With  $\epsilon_{\rm w} =0$, the model 
matches the observed distribution (the left panel) qualitatively. 
To understand how the bimodal distribution is produced, let us have a 
close look at how quenching proceeds in this model.  
Based on the model described in \S\ref{sec_rampressure}, 
a complete stripping of the gas disk occurs when the density of the 
surrounding medium reaches a critical value,
which can be obtained from Eq.\,(\ref{ram_pressure_0}) by
setting $R_{\rm rp} = 0$,
\begin{equation}
 \label{rho_c}
 \rho_{\rm X,c} \sim \frac{2\pi G \Sigma_{\star,o}\Sigma_{\rm g,o}}{V_{\rm vir}^2},
\end{equation}
where we have replaced the speed of the satellite, $v_{\rm s}$, 
by the virial speed $V_{\rm vir}$ of the host halo. Only galaxies 
that have reached such a high density region can be completely quenched.
Figure~\ref{trajectory} shows the evolution of satellites
on three typical orbits in a host halo with mass $10^{13}\Msun$: 
a nearly circular orbit (blue),  a nearly radial orbit (red) and an 
elliptical orbit in the middle (green). Although it has been 
accreted for about $6$ Gyrs, the satellite on the circular orbit is 
still actively forming stars at present day. In contrast, the satellite 
on the radial orbit has almost completely lost its gas during the 
first approach to the peri-center, which happened less than $1$ Gyr 
after accretion. 

\begin{figure*}
 \centering
 \includegraphics[width=0.99\linewidth]{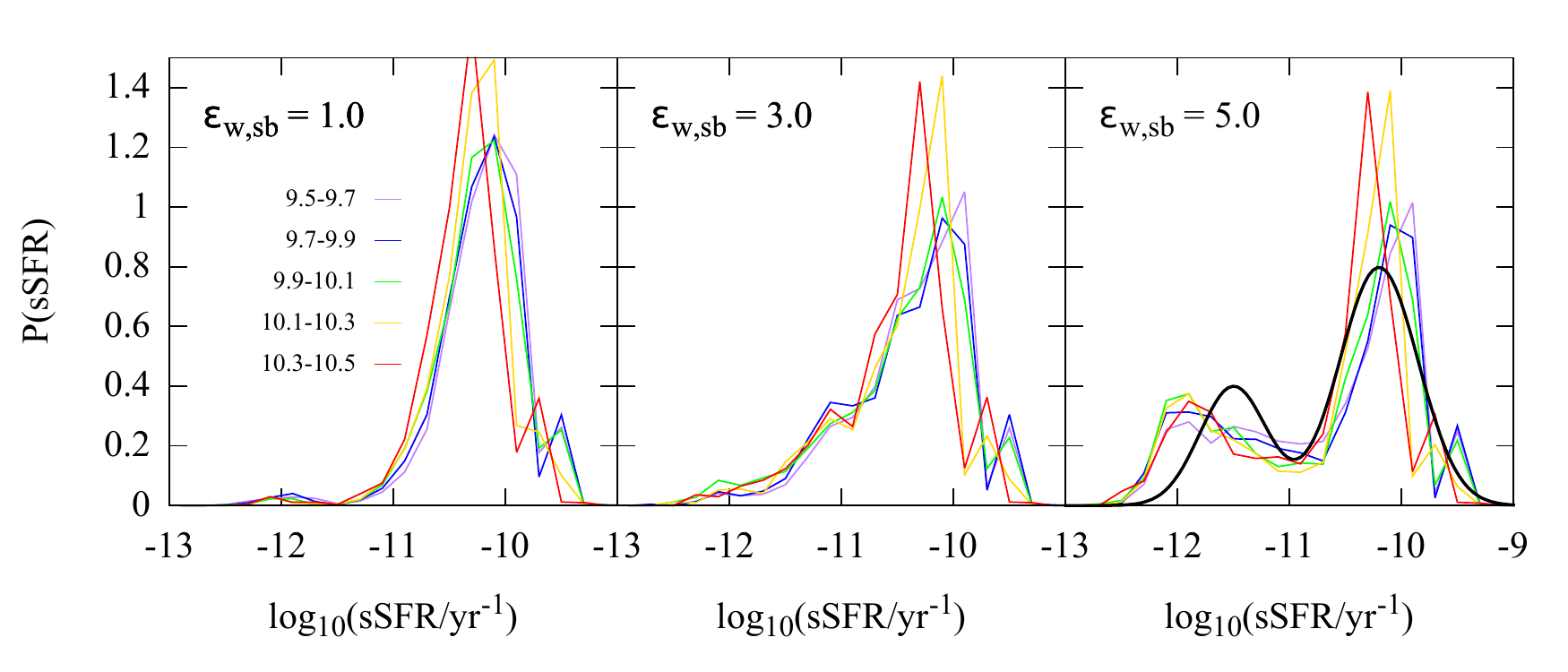}
 \caption{The distribution of the sSFR of satellite galaxies
          predicted by the model of tidally-triggered starburst. 
          Different panels correspond to different outflow mass loading 
          factors assumed for the starburst.
          The stellar mass range of satellites is coded in different colors, as
          indicated in the left panel. The black solid line in the last
          panel is the observational distribution 
          inferred from \citet{Wetzel12,Wetzel13}.}
\label{sSFR_KS_SB}
\end{figure*}

\begin{figure*}
 \centering
 \includegraphics[width=0.5\linewidth]{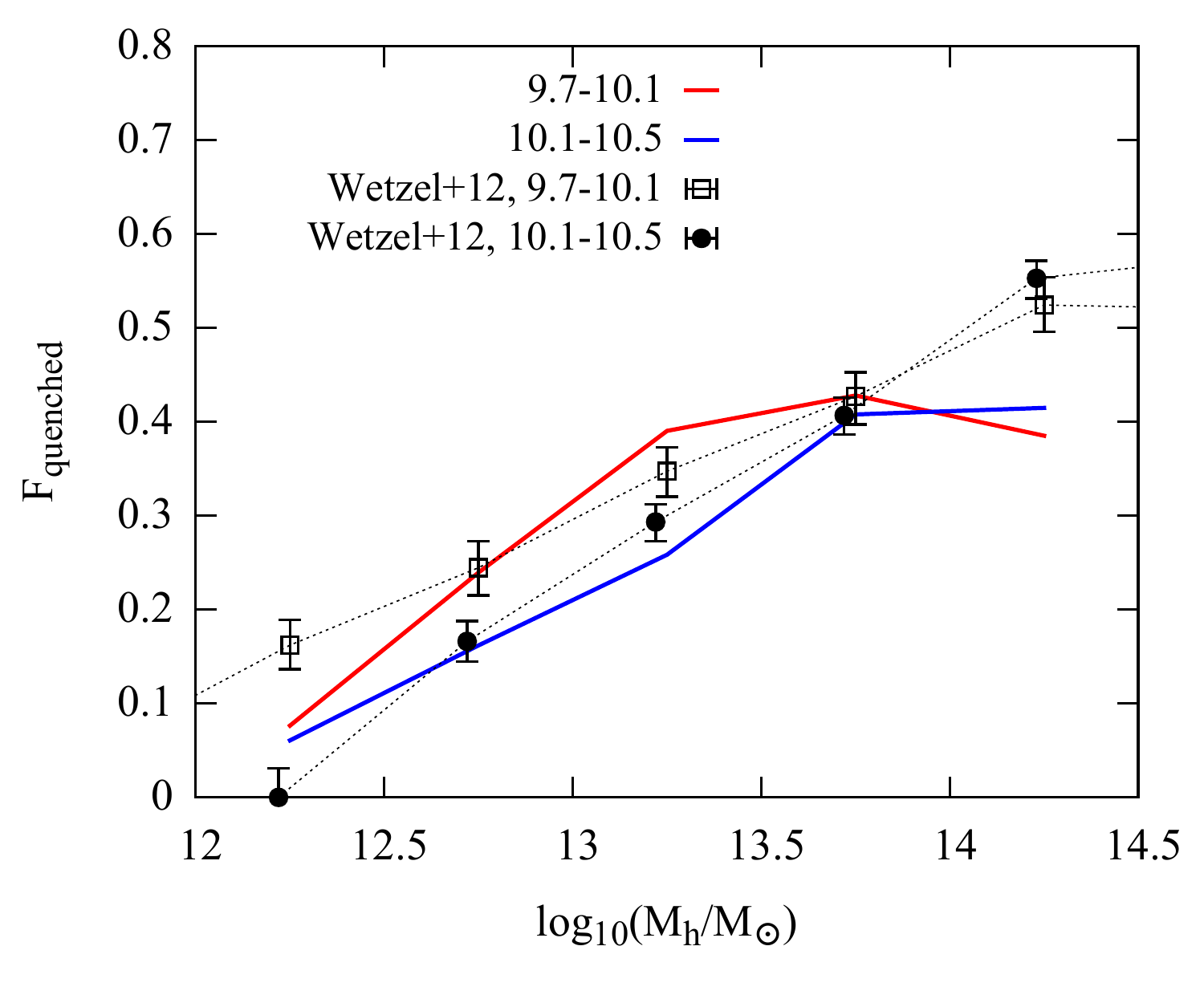}
  \caption{The fraction of quenched satellites as a function host halo mass.
          The colored solid lines are predictions by the model of 
          tidally-triggered starburst with $\epsilon_{\rm sb, w} = 5$.
          Data points with error bars are taken from \citet{Wetzel12}.}
\label{Fq_Mh_SB}
\end{figure*}

Compared with observational results, the predicted 
quenched fraction of satellites by ram-pressure stripping alone 
has much stronger dependence on host halo mass, quite 
independent of the assumed loading factor, as shown in 
Figure~\ref{Fq_Mh_RPon}.  There is a transition at 
a halo mass about $2\times10^{13}\Msun$ (for $\epsilon_{\rm} = 0$).
In more massive halos satellites are slightly over-quenched 
(by about $10\%$), while in halos with masses less than 
$5\times 10^{12} \Msun$ ram pressure by the halo gas 
seems unable to quench any satellites. This strong dependence 
on host halo mass is also responsible for the bimodal 
distribution seen in Figure~\ref{sSFR_KS_RPon}:
most of the star-forming satellites are from low-mass halos
($<10^{13}\Msun$) while most the quenched satellites are from
more massive halos ($>5\times10^{13}\Msun$). 

A simple analysis can be used to understand this strong dependence 
on halo mass. The critical density for complete stripping given by
Eq.\,(\ref{rho_c}) is proportional to $\rho_{\rm X, c} \propto V_{\rm vir}^{-2}$. 
Assuming an isothermal sphere gas distribution for the host halo, 
this critical density corresponds to a critical radius,
\begin{equation}
 x_{c} \equiv \frac{r_{\rm c}}{R_{\rm vir}} \propto V_{\rm vir}\,.
\end{equation}
It is this dependence on the virial velocity that produces 
the strong dependence of the quenched fraction on the host halo 
mass shown in Figure~\ref{Fq_Mh_RPon}.

\subsection{Quenching by Tidally Triggered Starburst}

Next, we switch off ram pressure stripping and investigate how 
the tidally-triggered star burst shapes the population of the
present day satellites. We can obtain a rough estimate of $x_{\rm sb}$
from \citet{Li08}. For an exponential disk $r_{90} \approx 4R_{\star}$. 
As shown by \citet{Li08}, significant enhancement occurs 
at a radius smaller than $\sim 5 r_{90}$, which leads to $x_{\rm sb}\approx 20$.
For simplicity we set $\epsilon_{\rm w}$ to $0$ and 
$\mathcal{E}_{\rm enhance}$ to $3$ and only tune the value of
$\epsilon_{\rm w,sb}$ to investigate the behavior of this model.

\begin{figure*}
 \centering
 \includegraphics[width=0.85\linewidth]{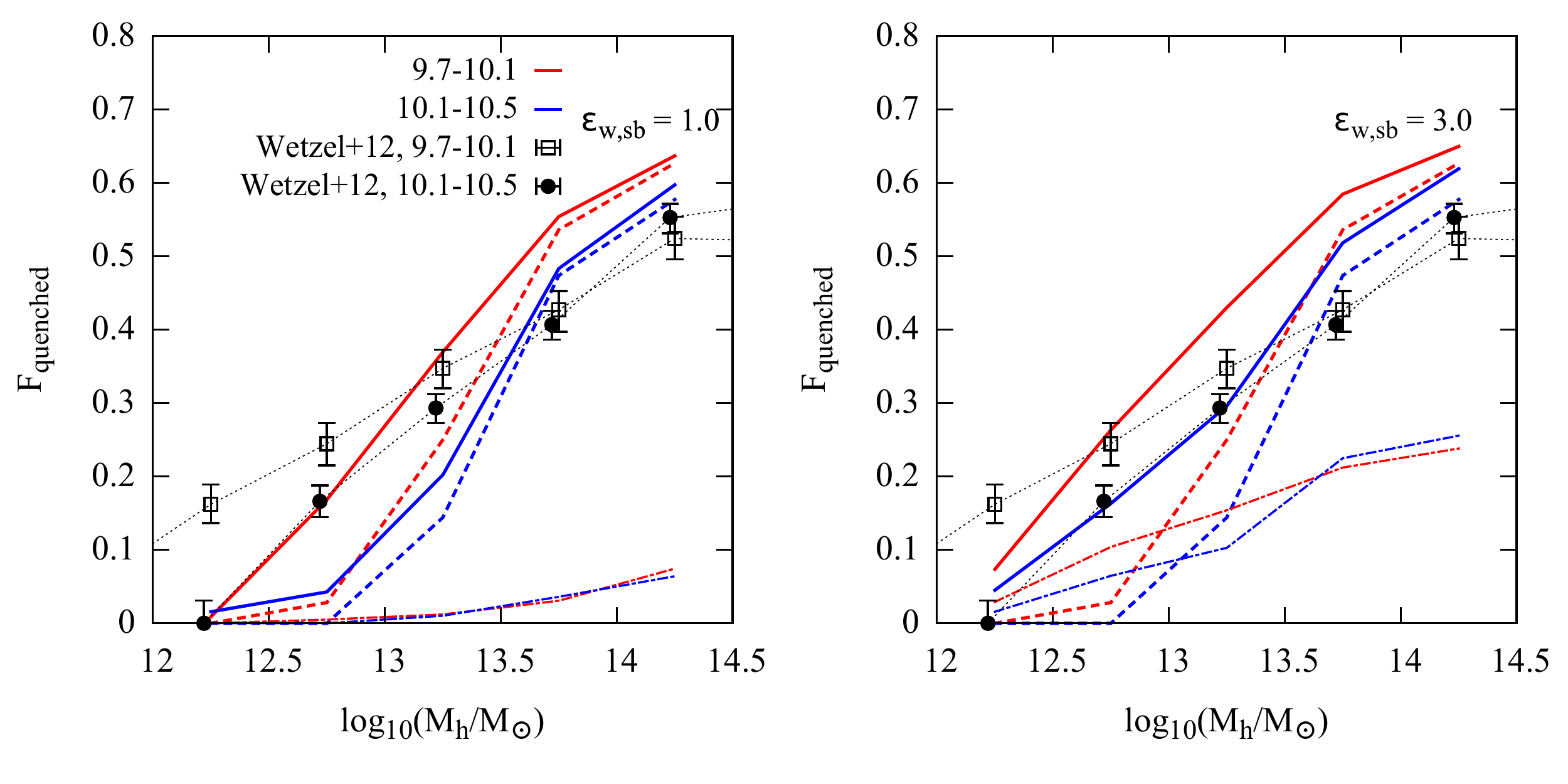}
 \caption{The fraction of quenched satellites as a function host halo mass.
          The dashed, dotted and solid lines correspond to models
          with only ram pressure, with only tidally triggered starburst,
          and with both processes, respectively.
          Left and right panels assume $\epsilon_{\rm sb, w}=1$ 
          and 3, respectively.} 
 \label{Fq_Mh_SB_RP}
\end{figure*}

\begin{figure}
 \centering
 \includegraphics[width=0.8\linewidth]{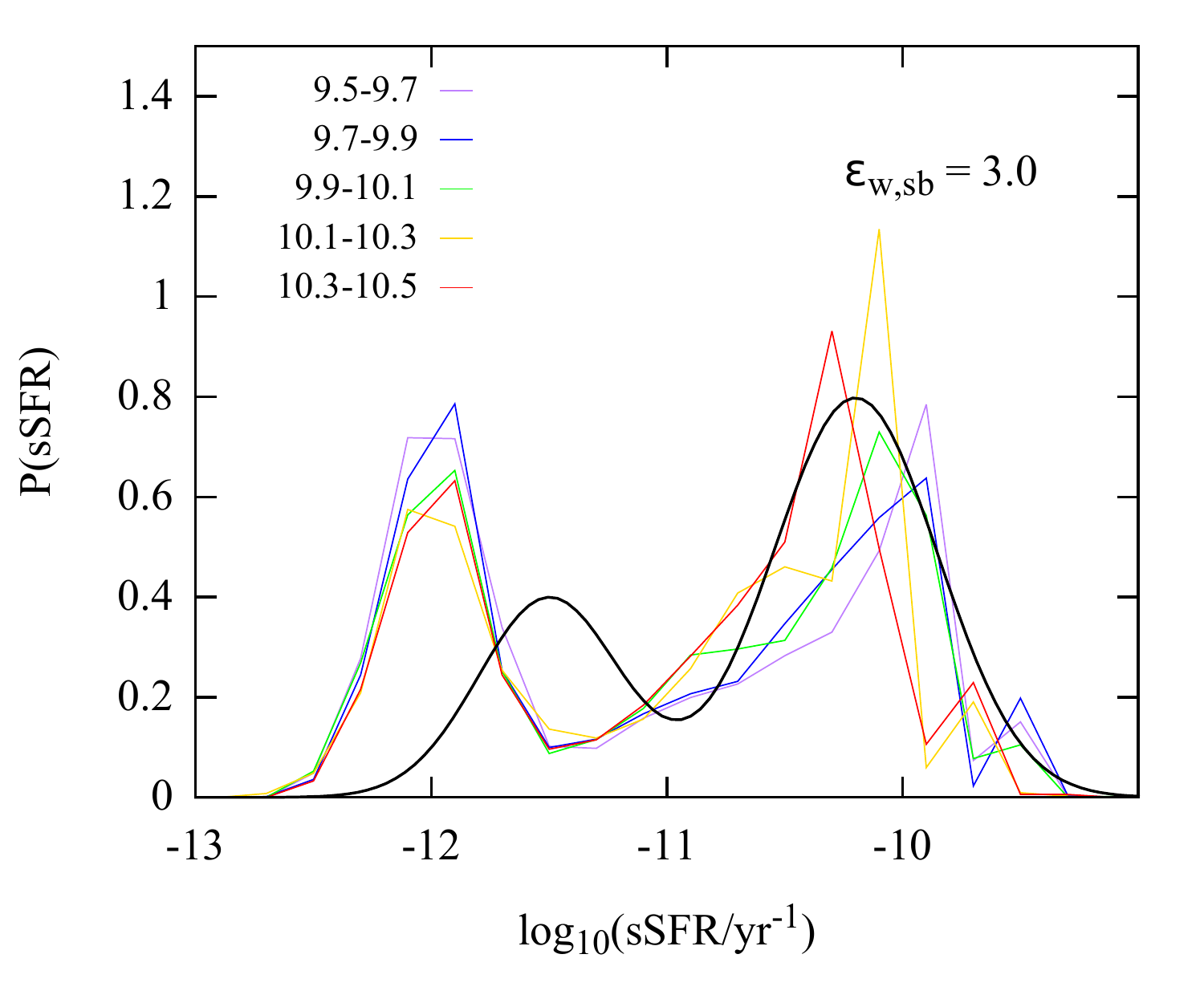}
  \caption{The distribution of sSFR predicted by the model in which 
           both ram-pressure stripping and tidally-triggered starburst 
          (with $\epsilon_{\rm sb, w}=3$)  are included.}
\label{sSFR_KS_SB_RP}
\end{figure}

\begin{figure}
 \centering
 \includegraphics[width=0.8\linewidth]{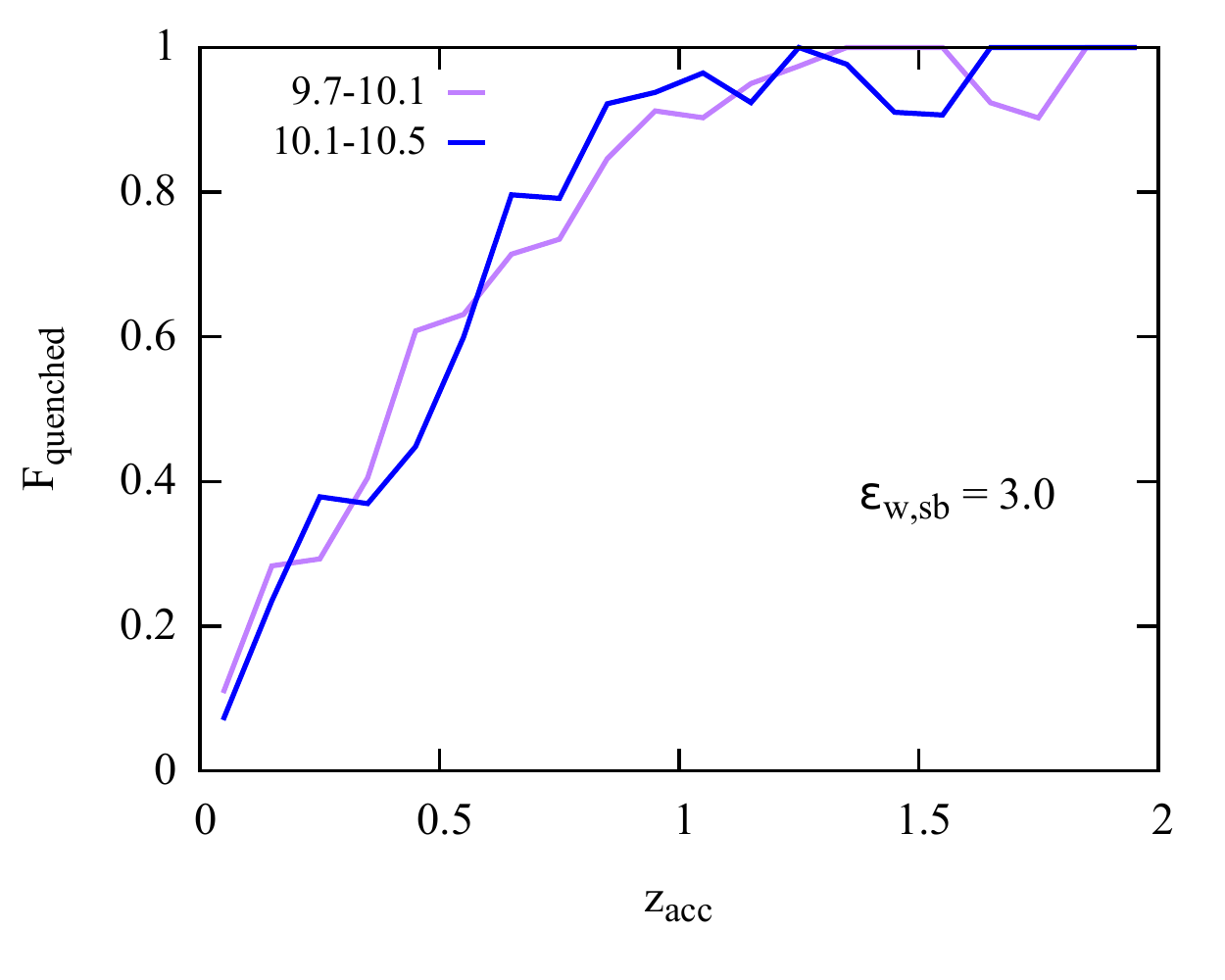}
 \caption{The fraction of quenched satellites as a function of the redshift
          of accretion, predicted by the model in which both ram-pressure 
          stripping and tidally-triggered starburst 
          (with $\epsilon_{\rm sb, w}=3$)  are included.}
 \label{Fq_zacc_SB_RP}
\end{figure}

The overall quenched fractions as functions of stellar mass
are shown in Figure~\ref{Fq_Msat_SB}, and the distributions 
of sSFR are shown in Figure~\ref{sSFR_KS_SB}.
A modest outflow during the burst phase ($\epsilon_{\rm w, sb}\le 3$)
produces a long tail in the quenched side, but only a limited number
of galaxies are quenched after the star burst. 
A significant bimodal distribution in the sSFR and 
a reasonable overall quenched fraction can be produced
if $\epsilon_{\rm w,sb} \approx 5$ is assumed.
This can be understood as follows.
The time for a satellite to cross the region within  
$R_{\rm sb}$ is about a few  $10^{8}{\rm yr}$, 
which is also the typical starburst time scale assumed 
in our model, while 
the gas depletion time scale by star formation, 
$\tau_{\rm dep} \equiv M_{\rm g}/{\rm SFR}$, is an 
order of magnitude longer. Thus, star formation alone 
can hardly consume a significant amount of gas during 
the star burst. Indeed, as shown in \citet{DiMatteo07},  
the extra gas consumed by star burst is not significant.
In order to get a complete quenching,
$\mathcal{E}_{\rm enhance}\left(1-\mathcal{R}+\epsilon_{\rm w,sb}\right) > 10$.

In contrast to the ram pressure stripping model, 
the quenched fraction predicted by the tidally-induced 
starburst model with $\epsilon_{\rm w, sb} = 5$
has a much weaker dependence on the host halo mass 
(Fig.\,\ref{Fq_Mh_SB}). The reason is that
the size of a central galaxy scales roughly with the 
virial radius of the host halos \citep{Mo98}.
If the starburst (including the outflow driven by the burst) is
strong enough to quench a flyby satellite in a short time scale, 
the dimensionless critical radius for satellite quenching
$x_{\rm c} \approx R_{\rm sb}/R_{\rm vir}$ is nearly
a constant.

\subsection{Conspiracy between Ram-Pressure Stripping and Tidally Triggered Starburst}

The existence of both the ram pressure stripping and 
tidally-triggered starburst have observational supports.  
Thus, both processes should be taken into account in 
a realistic model.  The two processes do not act independently. 
Here we demonstrate how they conspire 
to quench satellites in different environments.  
Figure~\ref{Fq_Mh_SB_RP} shows the predictions 
of two models, one with $\epsilon_{\rm w, sb} = 1$ 
and the other with $\epsilon_{\rm w, sb} = 3$, 
both including the joint effects of ram pressure stripping 
and tidally-triggered starburst.  For comparison, 
the predictions of models including only ram pressure or
only starburst are shown as dashed lines and dot-dashed lines, 
respectively. 
The sSFR distribution is shown in Figure\,\ref{sSFR_KS_SB_RP}.
As one can see, in halos with masses below
$10^{13}\Msun$, starburst with a modest outflow 
($\epsilon_{\rm w, sb} = 3$) alone
can hardly quench any satellites. 
Ram pressure stripping acting alone is also 
too weak to strip the gas discs in such halos, as already 
discussed in \S\ref{ssec_rampressure}. However, when 
both processes are turned on, the starburst can reduce the gas 
surface density significantly so as to make the ram pressure 
stripping more effective. In contrast, in more massive halos,  
ram pressure starts to work long before the starburst is 
triggered, so that the inclusion of starburst has only a modest 
effect. Overall the prediction of the combined model matches 
roughly the observed quenched fraction as a function of halo
mass and the sSFR distribution.
The largest discrepancy occurs for low-mass galaxies
in massive halos, where the model over-predict the quenched 
fraction by about $30\%$.  For low-mass galaxies, the 
predicted halo mass dependence of the quenched fraction 
also appears to be stronger than in the data.  Therefore, some halo 
mass-dependent prescriptions seem to be needed 
for the model to match the data better. For example, 
if the strength of tidally triggered starburst decreases 
with halo mass, or if the efficiency of ram pressure stripping in 
massive halos is lower than that given by the model, the match between 
model and data can be improved. The first possibility is actually 
likely.  In halos with masses above $10^{13.5}\Msun$,  ram pressure 
stripping starts to work long before the starburst is triggered. By 
the time when a satellite galaxy reaches $R_{\rm sb}$, 
the cold gas mass in its disk has already been reduced 
by a large amount, and the interaction of the gas-poorer 
disk may not be able to trigger a strong starburst. In the extreme case 
where no starburst is triggered for satellites in these massive 
halos, the quenched fractions would be reduced to those represented 
by the dashed lines in Figure~\ref{Fq_Mh_SB_RP}, which match
the data better. This possibility is not in conflict with the 
result of  \citet{Li08}, because the result is for the star-forming 
population, which is dominated by galaxies in relatively low-mass 
halos.

The wind loading factor, $\epsilon_{\rm w, sb} = 3$, required for 
low-mass halos appears to be reasonable for starburst galaxies. 
Observations of galactic winds show that the mass in the outflow 
gas is typically a few times the star formation rate for starburst 
galaxies \citep[e.g.][]{Martin99}. Once the gas is ejected from a satellite 
and becomes diffuse, ram-pressure and tidal force may strip it from the sub-halo 
of the satellite, even if the wind is not energetic enough   
to escape the sub-halo by itself.     

The quenched fraction is a strong function of the accretion time of 
the satellites (Figure\,\ref{Fq_zacc_SB_RP}). It increases rapidly
from $0.5$ at $z\sim 0.5$ (corresponding to a look-back time
of $\sim 5$ Gyr) to $1$ at $z\sim 1$ (corresponding to a look-back time of 
$7.5$ Gyr). This suggests that a satellite accreted at $z > 0.5$ is more 
likely to have already been quenched, while a satellite accreted at a later 
time is more likely to be observed as a star forming-galaxy.  
This is consistent with the finding of \citet{Wetzel13}, 
which shows that on average a galaxy
can remain as star forming for $\approx 4\,{\rm Gyr}$ after 
accretion before it is quenched abruptly.

\section{Summary}
\label{sec_summary}

We investigated the star formation quenching of satellite galaxies 
using a set of simplified but physically motivated models. 
The models are built upon the dark halo merger histories of
\citet{Parkinson08} and the sub-halo model of \citet{Taylor01, Taylor04}
that traces the the orbits, tidal stripping and disruption of sub-halos.
Instead of following the whole histories of present-day satellites,
we only keep track of the evolution of a satellite after it is accreted 
into a more massive system. The progenitors of satellites at the time of 
accretion, which are themselves central galaxies at that time, 
are initialized according to the dependences of stellar mass, star 
formation and cold gas content of star forming galaxies on 
halo mass as given in \citet{LZ15b}. Such dependences are 
constrained using current observations of the galaxy populations
at different redshifts. This approach avoids some of the serious 
uncertainties in the assumptions made in previous models of 
satellite galaxies, and allows us to focus on the quenching of star formation 
by intra-halo environments and processes. Our model does not 
assume a diffuse gas reservoir in sub-halos, and as such  
it assumes instantaneous ``strangulation".

We tested a series of models against the observational constraints 
derived from galaxy groups by \citet{Wetzel12, Wetzel13},
including the sSFR distribution, its dependence on stellar 
mass of the satellite and on the mass of the halo in which the satellite 
resides. We found that these observational constraints provide 
important information regarding the process of environmental quenching.

We first considered a model in which the satellites are quenched
by gradual gas consumption due to star formation and associated outflow.
Without any outflow, all the survived satellites would lie in the star forming
sequence. This means that the cold gas reservoir brought in at the time of 
accretion is large enough to sustain the star formation in satellites
all the way to the present-day.
A mass loading factor larger than unity produces an over-quenched 
satellite population. This critical loading factor is modest compared
with what is traditionally assumed in SAMs of galaxy 
formation. The over-quenching of satellites in these SAMs has 
usually been attributed to the instantaneous ``strangulation", and 
solutions to it have been focused on slowing down the stripping of the 
diffuse gas from sub-halos \citep{Kang08,Weinmann10}.
For instance, \citet{Weinmann10} suggested that no ram pressure of diffuse
gas should happen in order to reproduce the correct quench fraction,
in conflict with simple physical expectations. An alternative solution
to this over-quenching problem is to assume a modest outflow, 
as is demonstrated by our calculation and supported by other studies 
of central galaxies \citep{Lu14, LZ15b}. Thus, we believe that the 
so called ``over-quenching" problem is primarily due to the assumption 
of a high loading factor in these models. However, gradual gas 
consumption by star formation and outflow alone cannot produce 
a bimodal distribution for sSFR, suggesting that additional quenching 
mechanisms are needed.

The second model we tested is the ram pressure stripping of the cold gas
by hot halo gas. With realistic assumptions about the hot gas profiles in dark matter halos, 
quenching of star formation 
is found to occur in an abrupt manner when
satellites reach deep enough into the halo center, where the ram pressure
is sufficiently large to completely strip the cold gas. 
To obtain an overall quenched fraction that matches observation, 
this model does not need any net outflow powered by star formation. 
Ram pressure stripping is also able to reproduce a bimodal
distribution in the sSFR.  However, this bimodality is largely due to 
the strong dependence of the ram pressure stripping on halo mass: 
in massive halos ($>10^{13.5}\Msun$), the satellite population 
is slightly over-quenched compared with observation, 
while in low mass groups ($<10^{13}\Msun$),
ram pressure is too weak to produce any quenched satellites.
This is in conflict with the observational data in which a significant 
fraction of satellites in low-mass halos are also quenched. 

In a third model, we proposed a phenomenological 
prescription to describe starbursts that are triggered by tidal interactions
of satellites with central galaxies as they pass by the halo 
center. This model is motivated by and calibrated with   
the observational results of \citet{Li08}. The SFR
is assumed to be enhanced by a constant factor 
$\mathcal{E}_{\rm enhance} = 3$ (1 means no enhancement)
once a satellite reaches
a critical radius $r_{\rm sb} \sim 20 R_{\star}$. 
The enhanced star formation alone cannot quench the flyby satellites,
as the crossing time scale of the sphere $r_{\rm sb}$ is too short.
Without ram pressure, a strong outflow with a loading factor 
$\epsilon_{\rm w, sb} > 5$ during the starburst phase is required.
The dependence of the quenched fraction on halo mass 
in this model is found to be much weaker than the prediction of 
the ram-pressure stripping model. In the
halo mass range $<5\times10^{13}\Msun$, the prediction matches
observation, while in more massive halos the 
critical radius is too small to produce a large enough population 
of quenched satellites to match observation. 

Finally, we considered a model where both ram-pressure stripping and 
tidally-triggered starburst operate together.  With ram pressure, 
the wind loading factor in the starburst required to match observation 
is reduced. In low mass halos, ram pressure and a starburst with 
$\epsilon_{\rm w,sb} \approx 3$ working together can produce a 
reasonable match to the observed quenched fraction. In high mass 
halos, quenching is always dominated by ram pressure. 

In conclusion, our investigation demonstrates that, once the progenitors of 
satellites are properly modeled, the star formation properties of the satellite 
population can be understood in terms of physical processes that are expected 
to operate in galaxy groups/clusters. Further investigations are clearly needed. 
The assumptions regarding ram pressure stripping and tidally induced 
starburst need to be tested with numerical simulations. 
Detailed model predictions can be made by controlled simulations 
which follow the evolutions of satellites in different halos
with properly-set initial conditions. 
Further observational consequences, such as spatial 
distributions and velocity dispersions of quenched and 
non-quenched satellites in their host dark matter halos, should be explored 
to test the model in more detail. We will come back to some of these 
problems in the future.      

\section*{Acknowledgements}

We thank Yu Lu and Frank van den Bosch for helpful discussions.
HJM would like to acknowledge the support of NSF AST-1109354.

\end{document}